\documentclass[10pt,letterpaper]{article}
\usepackage[top=0.85in,left=2.75in,footskip=0.75in,marginparwidth=1in,margin=1.0in]{geometry}
\usepackage[utf8]{inputenc} % set input encoding (not needed with XeLaTeX)
\usepackage{graphicx} % support the \includegraphics command and options
\usepackage{cite}% clean citations
\usepackage{physics}%bra's and kets
\usepackage{amsmath}% Matrices
\usepackage{gensymb}%Degree symbols etc

\usepackage{booktabs} % for much better looking tables
\usepackage{array} % for better arrays (eg matrices) in maths
\usepackage{paralist} % very flexible & customisable lists (eg. enumerate/itemize, etc.)
\usepackage{verbatim} % adds environment for commenting out blocks of text & for better verbatim
\usepackage{subfig} % make it possible to include more than one captioned figure/table in a single float
\usepackage[blocks]{authblk} %for authorship
\usepackage{graphicx}%for images
\usepackage{float}%for image placement

\graphicspath{ {LaTexFigures/} }

\newcommand*{\vcenteredhbox}[1]{\begingroup
\setbox0=\hbox{#1}\parbox{\wd0}{\box0}\endgroup}

\usepackage{fancyhdr} % This should be set AFTER setting up the page geometry
\usepackage{sectsty}
\allsectionsfont{\sffamily\mdseries\upshape} % (See the fntguide.pdf for font help)
\title{Mode Characterization for Planar and Corrugated Multilayer Structures via Scattering Matrix Analysis}

\author{Casey Kneale\thanks{ckneale@udel.edu}\,\,}
\author{Karl S. Booksh }
\affil{Department of Chemistry and Biochemistry \\
University of Delaware, Newark, Delaware 19716 \\
}

\author{Jean Chandezon\thanks{jean.chandezon@univ-bpclermont.fr} }
\affil{Department of Physics \\
Université Blaise Pascal, Institut Pascal,
4 Avenue Blaise Pascal, TSA 60026, CS 60026, F- 63178 Aubière CEDEX
CNRS, UMR 6602\\
}

\begin{document}
\maketitle

\section*{Abstract}
The construction of the unconditionally stable planar rank 2 scattering (S) matrix for stratified systems is detailed from Fresnel equations. Several matrix decompositions and numerical calculations performed on the planar S matrix allow for the expedient characterization of purely absorbing, brewster, surface plasmon, and wave-guide modes. A figure of merit is presented from the decompositions of the scattering matrix constructed from the Chandezon method for corrugated surfaces. This figure of merit represents the hyper-area of the scattering matrix transform and allows for rapid distinguishability between lossy absorption phenomena, and surface plasmons. Some extension of this technique is possible for surface plasmon polaritons in the infrared region.

% the * after section prevents numbering
\section*{Introduction}
An upsurge of interest has been directed towards the discovery of novel plasmonic materials for applications which range from telecommunications to surface enhanced Raman spectroscopy. Most of the widely used theoretical tools for assessing lossy or interference behavior in multilayered structures are limited to computationally intensive simulations of electric or magnetic fields through stratified layers. These methods include: finite element method(FEM), finite difference time domain analysis(FDTD), and field expansion methods. All of these methods require further assessment of the calculated fields or transient responses to discern what type of phenomena(resonances, absorption, etc) are present. Due to the extra analysis’ such techniques can be seen as inefficient for large-scale material optimization problems because of memory and storage requirements. However, some work has been done on the analysis of system scattering matrices for the presence of modes on staircase approximated surfaces.

It has been well established that waveguide and resonant modes can be characterized via the presence of poles in the scattering matrix(S-matrix) at a given coupling condition \cite{NumMethods, ResPoles, Roundtrip}. Two commonly employed pole finding techniques discern what coupling conditions in a system cause either the  determinant or  maximum modulus eigenvalue of the inverse S-matrix to approach zero. These techniques have been successfully employed for certain numerical methods, namely the fourier modal method(RCWA). The determinant method requires a sufficiently large S-matrix, met with the paradox that such a large matrix becomes numerically unstable or cumbersome on inversion. More recently, schemes which linearize the S-matrix through its maximum/minimum modulus eigenvalues and search for roots in the solution space with Newton-Raphson or Halley algorithms have been investigated\cite{NumMethods}. None of the aforementioned methods, despite being rigorously sound, offer a means to differentiate the nature of the modes which are found as poles or other-wise.

In this manuscript we present novel adaptations to previously known methods for general mode identification in small S-matrices for both planar(Part I) and boundary matched periodic stratified structures(Part II) via a spectral approach. The rank 2 S-matrix for planar media is derived from the computationally inexpensive Fresnel equations. A simple scheme of physically guided decompositions of the planar S-matrix allows for the distinction between energy absorbing modes such as surface plasmons, purely absorbing modes, brewster modes, and waveguide modes in the case of total internal reflection. The periodic corrugated structures presented here-in also feature small(relative to other techniques) S-matrices due to the elegant nature of the Chandezon method \cite{RCWAvsCmeth}. With decompositions of the corrugated S-matrix we can discern between classic surface plasmons, absorbing modes, and in some cases so called surface plasmon polaritons. Some advantages to these approaches, caveats, potential applications, and notes on implementation are also featured.

\section*{Definitions}
Several esoteric conditions and mathematical operations are utilized on many occasions here-in. For brevity and reader clarity these have been abbreviated as follows,
\begin{itemize}
  \item "Conservation of Energy", in this paper represents the sum of the power for all reflected or transmitted modes from an optical element. In a lossless medium this is 1., but for dielectric materials it is typically $<$ 1.0 depending on physical conditions.
  \item SP, Surface plasmon, a plasmonic resonance provided $\Re ( \epsilon ) < 0$ and  $\Im ( \epsilon ) < 0$
  \item SPP, Surface plasmon polariton, a plasmonic resonance provided $\Re ( \epsilon ) > 0$ and  $\Im ( \epsilon ) > \Re ( \epsilon )$
  \item DI, determinant inverse, $DI = \frac{1}  {| \hat { S }| }  = \frac {1} {\prod eig( \hat { S } )} $
  \item MMEV, Maximum modulus eigenvalue inverse, $MMEV= \frac{1}  {Max[|EigVal(\hat { S } )|]} $
  \item RMSV, reciprocal maximum singular value, $RMSV = \frac {1} {Max[SV(\hat { S } )]}  $
  \item C-figure, product of singular values from the imaginary part of the truncated global S matrix, $C-figure = \frac{1} {\prod SV( \hat { \Im (S_{trunc}) })} $
\end{itemize}

\section{Planar Case}
\subsection{Fresnel Equations and Development of the Planar Scattering matrix}
Many schemes have been developed to model multilayer planar media with the Fresnel equations. Most of the implementations that have disseminated from the literature and associated texts utilize the 2x2 transfer matrix method(TMM). Although TMM offers similar reflection(R) and transmission(T) coefficients, it is inherently unstable. The instabilities arise from the products of large exponent which represent the phase of incoming radiation with-in layers. When a layer thickness approaches an order of magnitude of ~5-7 times greater then the wavelength of light, numerical errors can occur due to growing exponential terms(S1). In many physical situations TMM forces a practitioner to apply semi-infinite approximations and not model optically thick materials. Due to the aforementioned issues with TMM and that a similar construction of the S-matrix is used for modeling corrugated surfaces, the implementation of the unconditionally stable 2x2 scattering matrix for TM polarized light is briefly detailed.

The interfacial layer transfer matrix is first developed from the layer reflection and transmission coefficients,
$$I_{n,n-1} = \frac{ 1 } { t_{n,n-1} } \cdot 
\quad \begin{bmatrix}
	1 & r_{n,n-1} \\ r_{n,n-1} & 1
\end{bmatrix} \quad
$$

Where the interfacial scattering matrix\cite{LisSMatrix} can be defined as,
$$\widetilde s =
\quad \begin{bmatrix}
 	I_a - {I_c \cdot {I_d}^{-1} \cdot I_b} &  I_c \cdot {I_d}^{-1}  \\ 
	-I_b \cdot {I_d}^{-1} & {I_d}^{-1}
\end{bmatrix} \quad
$$
Where a,b, c, and d are the clockwise components of the interfacial layer transfer block matrix. The layer scattering matrix can be derived from the half diagonal airy-drude light attenuation coefficient($\phi$).
$$S  = 
\quad \begin{bmatrix}
 	1  & 0  \\
	0 & \phi  
\end{bmatrix} \quad \cdot \widetilde s \cdot 
\quad \begin{bmatrix} 
	\phi  & 0  \\
	0 & 1
\end{bmatrix} \quad =
\quad \begin{bmatrix}
	 \frac{\phi}  {t}(1 - r^2)  & r \cdot t \\
	 -\phi^2 \cdot r \cdot t & \phi \cdot t
\end{bmatrix} \quad
$$

Where $\phi$ represents the phase change of the radiation relative to the incident plane. It is defined by the wavelength($\lambda$), layer depth($d_n$), Snell's law transmitted angle ($\theta_N$), and refractive index of the layer($RI_n$).

$$\phi = \exp(2i \pi \lambda \cdot RI_n \cdot  d_n \cdot \theta_n)$$

The system scattering matrix($\hat S$ ) is then constructed via successive star products for each layer S matrix until N layers are described,

$$ S_{0} = I$$
$$\hat { S } = ( ( (S_{0} \star S_{1}) \star S_{2} ) ... \star S_{n})$$

Where the system reflection(R) coefficient, transmission(T) coefficient, and the figure of merit enhancement factor (E) can be found from simple manipulations of the S-matrix,

$$R = |\hat { S }_{2,1}|^2 \quad
T = \frac{\theta_N \cdot |\hat { S }_{1,1}|^2} { \theta_{0} } \quad
E = |\hat { S }_{1,1}|^2$$

Where $\theta_0$ is the incident angle. Please note this matrix designation is the transpose of the experimental convention described by Rumpf\cite{Rumpf}. The incoming field amplitudes can finally be related to the outgoing field amplitudes by the definition of the system scattering matrix which represents the entire planar system (Figure 1).
$$ \ket{ \psi_{out} }= \hat { S }   \ket{ \psi_{in}}$$
\begin{figure}[H]
	\includegraphics[width=6cm]{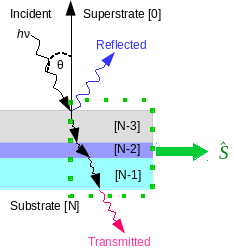}
\centering
	\caption{ A cartoon depicting a stratified system with boundary matched planar media. Layer and angle of incidence information are all stored in the global scattering matrix(green). }	
\centering
\end{figure}

\subsection{Planar Scattering Matrix Mode Spectral Analysis}
	The goal of the following scattering matrix analysis’ is to develop numerical methods which discern the character of energy absorbing modes present under given physical conditions. Very rarely do the determinant inverse(DI), maximum modulus eigenvalue(MMEV), or reciprocal maximum singular value (RMSV), of the system S matrix reach a root with-in machine tolerances in the planar case. Instead they present as minima due to the limitations of low rank. Many such extrema may be numerical artifacts or other nuances(critical angle, etc). Therefore, the extrema observed in such an analysis’ are not necessarily modes of interest and more than one criterion must be made for any such characterization. 

	The problem of mode characterization via the S matrix cannot be presented in a facile manner. For example, a seemingly reasonable criterion to test if a conservation of energy minima is a surface plasmon might be to confirm that either a pole is present in the S-matrix or a maxima in enhancement factor is found at the same spectral location. This was an unestablished but often true statement that many surface plasmon resonance spectroscopists have applied, perhaps inadvertently, for many years. However, both brewster, and total internel reflection(TIR) waveguide modes can all exist under these same criteria. Most notably SP resonances have been shown to occur at different coupling conditions than those of their respective conserved energy minima\cite{Nature}.

\subparagraph{Purely Absorbing Modes}
Foley et al demonstrated that in the Kreatschman-Raether configuration($\lambda$ = 532nm) for a stratified structure composed of glass, germanium(4nm), gold(50nm), and air the reflectance minima(54\degree) is observed 9\degree away from the known resonance condition(45\degree). They termed the mode(54\degree) which did not meet dispersion relations or propagate strongly as a purely absorbing(PA) mode. Simulations, performed under identical conditions, confirm that an enhancement factor(E, or $|S_{1,1}^2|$) maxima presents at 45.57\degree with a neighboring reflectance minima at 53.94\degree. The authors show through FDTD calculations and considerations from the modal equations that the true SP condition can be reasonably identified by the enhancement factor maxima. None of the matrix decompositions of the native S-matrix presented in this study afford identifiable features at the location of the true SP coupling. 

	By nulling either or both of the reflection ports ($S_{2,1}$, $S_{1,2}$)  of the S matrix we can find minima at the locations of the enhancement max for the true SP which corresponds to a PA mode via the MMEV approach. An example of the classic 4-layer BK7 glass, chromium, gold, water, Kreatschman-Raether configuration is given below(Figure 2). In the simplest case, that of nulled off-diagnol entries, the MMEV simply represents the reciprocal maximum diagonal transmission entry. Considering that these SP modes are found without conservation of energy minima these modes can be made distinct from other SP modes if the respective PA mode is also found. However, a different criterion must be made in order to identify whether a mode is of brewster nature or not.
\begin{figure}[H]
	\includegraphics[width=7cm]{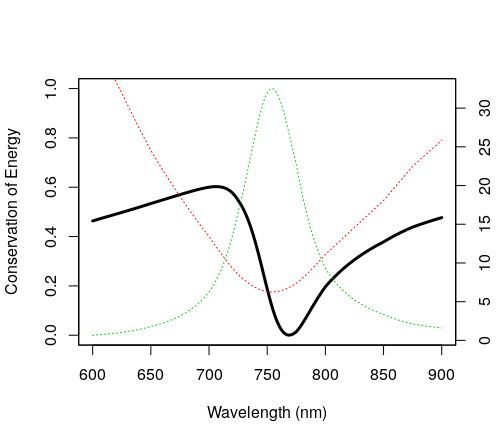}
\centering
	\caption{ Wavelength interrogation at an angle of incidence of a 67.0\degree for stratified structure which consists of BK7 glass, 5nm chromium, 50nm gold, semi-infinite water. The PA mode is located at the conservation of energy minima(black). 769nm, and the SP mode is located at the enhancement factor maxima(green)/nulled MMEV minima(red), 754nm. }	
\centering
\end{figure}

\subparagraph{Brewster Modes}
	Unlike PA modes, brewster modes, sometimes exist with a concomitant minima in conservation of energy provided that a material is lossy. Brewster modes can be defined as modes which have only incident and transmitted plane waves \cite{Book}. The DI can allow for the characterization of brewster modes by again nulling either reflection parameters. Without, reflection, the DI is simply the reciprocal product of the transmitting ports. At brewster modes, there is little energy distributed to backwards transmission and an increase in the reciprocal is observed. Thus a maxima is observed in the brewster coupling locations. Due to the similarity in the phenomena, the DI approach also yields maxima at SP critical angles. Modes such as classic SPs tend to have large backwards transmission and reflection parameters, such that a distinctions can be made. An example of a lossy substrate, polyaniline, exhibitting a brewster mode is provided(Figure 3). 
\begin{figure}[H]
	\includegraphics[width=7cm]{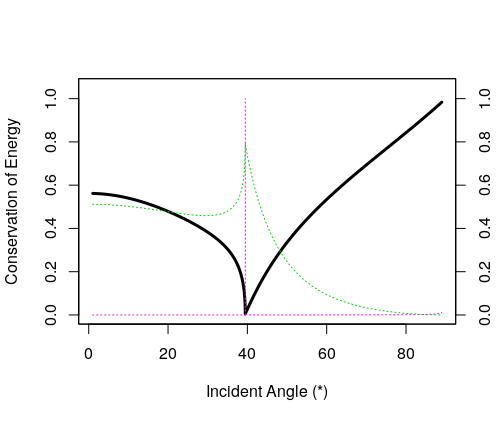}
\centering
	\caption{ Angular interrogation at an wavelength of 1.25um for a stratified structure which consists of polystyrene(RI = 1.573), 75nm polyaniline(RI = 0.652+0.351i), and semi-infinite vacuum. The brewster mode is evident at 39.47\degree by both the enhancement factor(green) and the scaled DI (pink) maxima’s. }	
\centering
\end{figure}

\subparagraph{Total Internal Reflection Waveguide Modes}
TIR wave-guide modes also feature enhancement maxima at their conservation of energy minima. Both the DI and MMEV methods tend to present sharp minima at these conditions. Thus a computationally demanding fine interval of spectral element(wavelength, or coupling angle) must be employed to assess such drastic changes. While the scheme of RMSV broadens these sharp changes by moving background effects to rotation matrices and affords minima at comparatively low spectral intervals. The RMSV has a resolution for finding such modes that is proportional to the conservation of energy interval spacing. Thus the use of RMSV seems optimal compared to the other techniques for efficient identification of wave-guide modes. 

	TIR coupled SPP’s have enhancement factor maxima that are shifted from their conservation of energy minima similar to PA modes. Unlike PA modes, none of the aforementioned numerical techniques afford minima or maxima near these TIR coupled SPPs. Thus far, only a heuristic approach for such a discrimination has been attained.

	A simple means to assess whether peaks presented in TIR conditions are due to SP modes is to use only the real permittivity in the construction of the layer S matrix for a potentially plasmonic material. This will remove minima in conservation of energy spectra attributed to the losses of the light on the material. The respective enhancement factor maxima will still remain in the spectra, albeit shifted. Any enhancement factor or conservation of energy minimum which is attributable to a classically plasmonic substrate will not be present. An example spectra in which all of the aforementioned phenomena is presented below(Figure 4). This is a heuristic test which suffers largely from the shifts in enhancement factor maxima incurred by changing the interfacial propagation angles in a medium.  
\begin{figure}[H]
	\includegraphics[width=7cm]{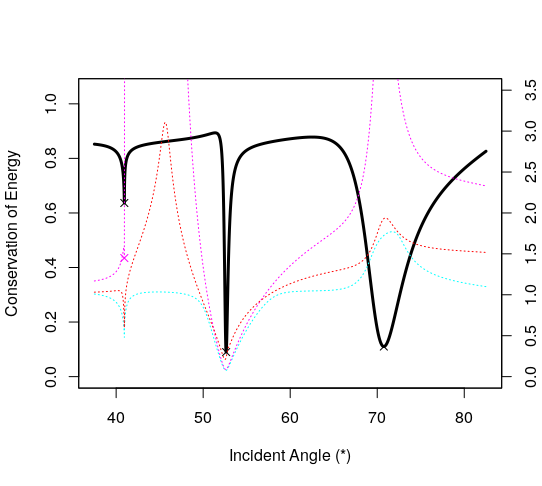}
	\includegraphics[width=7cm, height = 6.5cm]{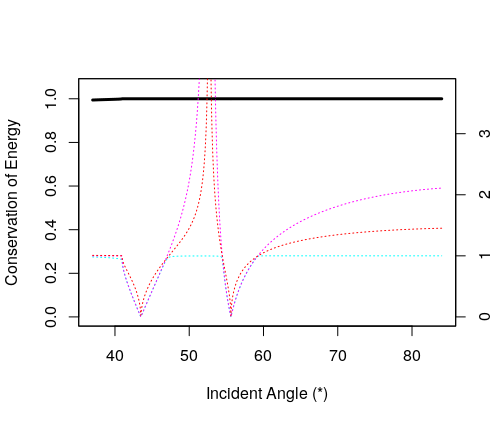}
\centering
	\caption{ Angular interrogation spectra of soda lime glass, silver(45nm), water(545nm), and vacuum with an incident wavelength of 535nm(left). The same stack is investigated with only the real permittivity of silver(right).  }	
\centering
\end{figure}

\subparagraph{Planar Mode Analysis Summary and Applications}
In summary, absorption phenomena can be distinguished from one another by their \textit{ conservation of energy} minima, enhancement factor maxima, and a few numerical techniques(Figure 5). It is left to the reader’s discretion to devise means to study their system S matrices for whatever their specialized problem may be. In order for the proposed schema to be deployed as a functioning algorithm some practical concerns must also be addressed.

The examples provided here-in are simple 3-4 layer arrangements. For more involved structures the presented numerical methods may need to be applied on local rather then system scattering matrices. If the incident media in a planar stack is lossy then shifts in the extrema of some of the numerical techniques can occur. Such that, some of the techniques described here-in may not hold in exactitude and thresholds may have to be set. Most practitioners of models which feature infinite half-spaces do not allow for lossy incident media for obvious reasons. Regardless, practical thresh-holds are advised for low resolution spectral interrogations, so that optimization procedures can be made efficient. Another simple means for improving computational efficiency is that of adaptive peak finding algorithms, which the authors recommend.
\begin{figure}[H]
	\includegraphics[width=10cm]{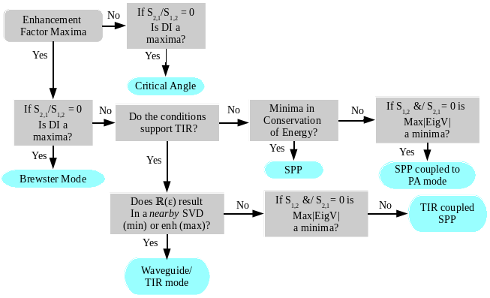}
\centering
	\caption{A proposed flow-chart for mode characterization based on the tests out-lined in preceding sections.  }
\centering
\end{figure}

\subparagraph{Dimension Reduction in Optimization Problems}
Consider that the modes in the planar system are a function of the dielectric functions of the materials, and coupling conditions(incident angle, wavelength). So far as radiation impinges upon or transmits through a given layer it can be seen that the S matrix possesses one more variable, layer depth. Thus, the S matrix contains information about the conditions for such modes almost irrespective of depth, provided the above statement holds true. Such that, if the goal of a spectral analysis is to find modes which meet a given criterion, the visibility of extrema in the conservation of energy spectra or enhancement factor is not necessarily a requirement. 

This can allow for the reduction of the depth dimension to optically relevent regions for optimization problems. An algorithm can be designed to explore material combinations at layer thicknesses above or below a materials penetration depth for certain modes. Then layer depth(s) can be optimized to satisfy a given constraint should a desired property be discovered. 

An example of such an experiment was performed for a semi-infinite fictitious substrate(RI = 1.35), thermally evaporated bismuth metal(thickness = 400nm, RI = 0.7773+12.80i), and semi-infinite vacuum with an incident wavelength of 3.1$\mu$m. An angular interrogation reveals that a mode is present at 48.03\degree and a brewster mode/critical angle at 47.99\degree while there is no such minima in the conservation of energy nor a maxima in enhancement factor(Figure 6). The thickness was crudely tuned to 55nm such that the signal to baseline ratio was maximized. This signal at 48.03\degree was confirmed to be plasmon-like via the RMSV minima and enhancement factor maxima. This result was also reproduced by field calculations. By classic dispersion rules, bismuth, should not be seen as plasmonic at this wavelength. Thus there appears to be applicability to surface plasmon polaritons (SPP) using the aforementioned criteria. It is interesting to note that when the depth approached 1nm that the mode at 47.99\degree behaved only as a brewsters mode by MMEV and field calculations. 
\begin{figure}[H]
	\includegraphics[width=7cm]{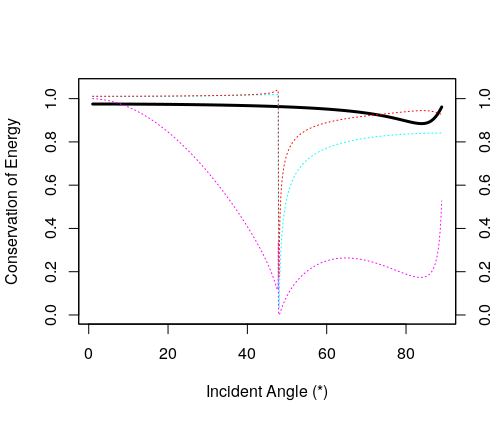}
	\includegraphics[width=7cm, height = 6.0cm]{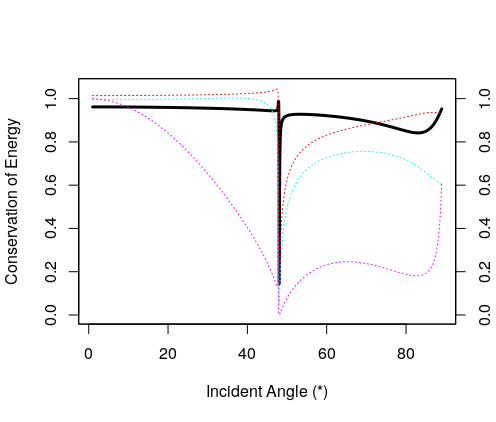}
\centering

	\caption{Conservation of energy spectra of 400nm Bi metal(left) and 55nm Bi metal(right). Both spectra have secondary axis for the following numerical methods: nulled DI (magenta), nulled MMEV(red), RMSV(cyan). }	
\centering
\end{figure}
Thus it can be seen that mode distinguishability in known and unknown materials can be performed. Perhaps more importantly, a continuum dimension in an optimization problem can be reduced to layer depth regions matching physical cases, IE: purely absorbing/reflection, transmitting/reflecting/absorbing, purely transmitting.
\section{Corrugated Case}
\subsection{The Chandezon Method and the Corrugated Scattering matrix}
The method chosen for constructing the scattering matrix for the stratified 1-D periodic corrugated structures(Figure 7) was the coordinate transform method discovered by Jean Chandezon. In the literature it is colloquioully known as the C-method, coordinate transform method, or the Chandezon method. A reader who is uninformed as to how the C-method solves Maxwell’s equations for periodic media is encouraged to read the original covariant formulation\cite{OrigCMeth}, the improved differential operator \cite{AdaptiveSpatial}, and the differential formalism\cite{DiffFormalism}. This method was chosen because it is computationally efficient relative to the fourier modal method(RCWA), features small scattering matrices, and represents periodic gratings in a natural manner. 
\begin{figure}[H]
	\vcenteredhbox{\includegraphics[width=7cm]{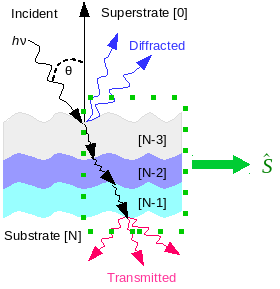}}
	\vcenteredhbox{\includegraphics[width=7cm]{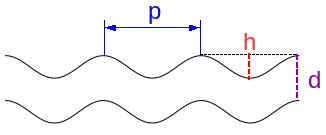}}
\centering

	\caption{A cartoon depicting the boundary matched periodic media in a stratified system(left). Layer and angle of incidence information are all stored in the global scattering matrix(green).  A graphic depicting the period(p), grating height(h), and layer depth(d) for the corrugated problem(right). }	
\centering
\end{figure}

The development of the scattering matrices follows the same equations as the planar case, however, a few adaptations are required due to the matrix size, the C-method’s depiction of mode attenuation, light polarization, and how the interfacial transfer matrices are formulated. The matrix size for the corrugated surfaces is technically infinite; due to the rules of fourrier factorization the matrix size is well approximated at finite truncation orders for most geometries. For details on the development of the scattering matrix for multilayered gratings see the the equations furnished by Lifeng Li in their seminal paper of multilayered structures using modal methods\cite{LisSMatrix} and the solution schema of Cotter et. al.\cite{Cotter}.

\subsection{Hypothesis of the Corrugated Scattering Matrix Transform}
The corrugated S-matrix presents special obstacles for mode analysis. For example, the MMEV, and DI present either minima or maxima in the presence of effectively any energy absorbing mode. Even in the instances of SPs calculated from elementary dispersion rules and finite element method. However, the approach presented by the RMSV is amenable to this architecture with some extension.

It has been well established that the scattering matrix maintains numerical stability by generating comparatively large components to balance the increasingly smaller ones when approaching the conditions for a resonant mode \cite{Roundtrip}. Our investigation showed that the first ket of the imaginary component S matrix was typically the loading most responsible for the variance of the principal components. Although, this ket is representative of the S matrix; consider it’s removal. The removal of this ket requires the removal of the corresponding bra in the incoming radiation condition. 

Thus the singular values of the truncated matrix represent scaling factors of a ‘scattering cross section’ (note: rather then the signed volume/determinant). The product of these scaling factors was then applied as a prototype figure of merit to describe a hyper-area of the scattering transform. From here on, this figure of merit will be referred to as the C-figure. This can be seen as similar if not equivalent to the treatment of the RMSV planar case. However, the preference for dimension removal has changed to reveal the S matrix unhindered by some of the artifacts that arise due to balancing large and small components.

\subsection{Corrugated Case Scattering Matrix Mode Analysis}
Simple test-cases were explored for proof of concept. Two substrates, gold and polyvinylpyrrilidone(PVP), were studied with respect to the magnitude of the C-figure and the effect of truncation order for a shallow sinuisoidal grating profile(Figure 8). The gratings had 1100nm periods, 50nm heights, and incident wavelengths of 550nm. The depth of the gold grating was set to 200nm and the PVP grating 400nm.

Exponential trends for the C-figure were observed at the locations for conservation of energy minima in PVP(mean R$^2$ = 0.9988 +/- 0.00042)  and locations of resonance for Au(R$^2$ = 0.9884 +/- 0.0037) across the truncation orders from 8-18. Most notably, the trend for absorption phenomena were exponentially growing from $10^{1}$ to $10^{9}$. While the trend for SP resonances was exponentially decaying from $10^{-2}$ to $10^{-6}$. These exponential trends were not found in conserved points attributable to numerical artifacts in randomly generated spectra.

\begin{figure}[H]
	\includegraphics[width=12cm]{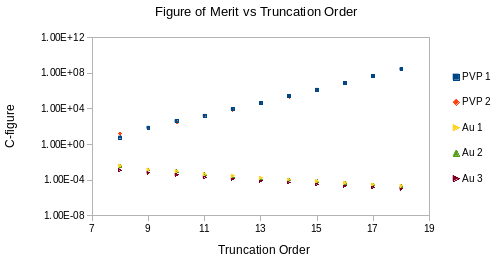}
\centering

	\caption{Semi-log plot which shows the exponential trends between the C-figure of conserved \textit{conservation of energy} minima for two periodic substrates amd truncation order. }	
\centering
\end{figure}

Another series of experiments tested both media at 30 randomly generated sinuisoidal profiles and wavelength conditions at sequential truncation intervals. Only minima that were conserved across all truncation orders which were not attributable to resonances, or meaningful absorption phenomena were studied. Of the retained points in the PVP spectra linearly increasing trends were observed(R$^2$ =  0.844 +/- 0.244). Similarly for the Au spectra a linearly decreasing trend was observed(R$^2$ = 0.898 +/- 0.253). For the samples which were deemed non-linear(R$^2$ $<$ mean), none followed exponential trends.From this small sample study a more wide-scale test for applicability was devised. 

Classically plasmonic substrates (Al, Au, Ag), absorbing substrates(PVP, water), and non-absorbing(silica glass, polystyrene) were assessed in 3 layer arrangements for modes they should not possess. The assessment criterion was as follows: if a C-figure of $>$ 1 was observed for a plasmonic material then a false positive was observed, and if a C-figure of $<$ 1 was observed for a non-plasmonic material then a false positive was observed. All 8 materials were tested for 250 random grating configurations/conditions, at a modest truncation interval(N=13) and no false positives were found. A thorough investigation for false negatives has not yet been performed due to the inherent complexity in implementing such a broad survey. However, the C-figure appears consistent across materials and configurations in discerning appropriate mode properties.

\subsection{C-figure Based Spectral Analysis}
The broad tests for the working principles of the hypothesis suggest that a binary pass/fail discrimination can be performed to see if an absorbing region is a classic plasmon or an absorbing mode. The figure of merit was applied spectrally in a similar manner as the numerical approaches demonstrated for the planar case. For SPs minima in total conservation of energy correlated to zeros(with-in tolerance/truncation order) or minima in the C-figure to be considered classically resonant SP modes(Figure 9). While other absorbing modes presented minima, but significantly larger in magnitude than zero. It is interesting to note that, the C-figure typically features singularities with respect to differentiation near SP modes but not absorbing modes. Although the C-figure does function to distinguish absorbtion from SP modes some discrepancies are observed for SPPs.
\begin{figure}[H]
	\includegraphics[width=7cm]{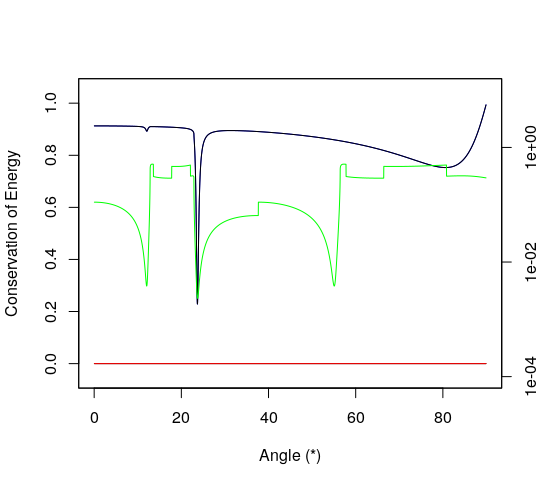}
	\includegraphics[width=7cm]{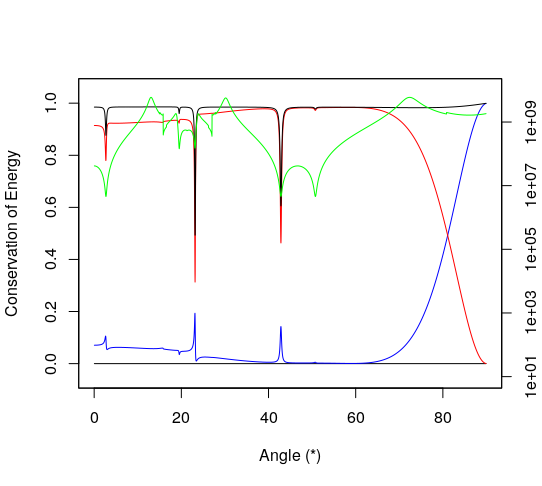}
\centering

	\caption{Conservation of energy spectra of an aluminum sinuisoidal grating(height = 25nm, period = 900nm, wavelength = 550nm) suspended in vacuum(left). Conservation of energy spectra of an PVP sinuisoidal grating(height = 75nm, period = 1100nm, wavelength = 800nm) suspended in vacuum. Reflection and transmission traces are shown in red and blue respectively. The secondary axis’ denote the magnitude of the C-figure (green).  }	
\centering
\end{figure}

When the C-figure has been applied to materials which have been established to exhibit SPP modes, conservation of energy minima do not coincide with the minima in the figure of merit space. For example, bismuth\cite{Bi} and antimony\cite{Sb} are known to host surface plasmon polaritons on trapezoidal grating profiles. Although, field calculations have shown that these SPP modes behave nearly identically to SP modes, there is a difference in their assessment by the C-figure. Instead of assessing the total conserved energy spectra, the correlation can only be seen between individual diffracted/transmitted order extrema and maxima in the C-figure space, and even then, quite loosely. Thus an assessment of the permittivity at a given incident radiation condition must be performed to assess this case. These maxima are typically have magnitudes of 100-101 and feature sharp peaks which require fine angular resolution($<$ 0.01\degree) to be observed. 

Good agreement for these resonances has been found for bismuth metal under experimentally determined conditions. Two first order SPP modes were found on Bi gratings(2000nm thick, 50\% duty cycle, trapezoidal, 20$\mu$m period, 1.4$\mu$m amplitude) at 31.70\degree(C-figure: 31.67\degree) and 64.16\degree(C-figure: 64.16\degree) with an incident wavelength of 9.5$\mu$m and a modest truncation order of 15. The percent differences between the location of the known SPPs in bismuth are $<$ 0.1\%. However, only a heuristic threshold can allow for such a characterization to be automated. Similar but less accurate results were obtained for known SPPs on an antimony grating.

\begin{figure}[H]
	\includegraphics[width=7cm]{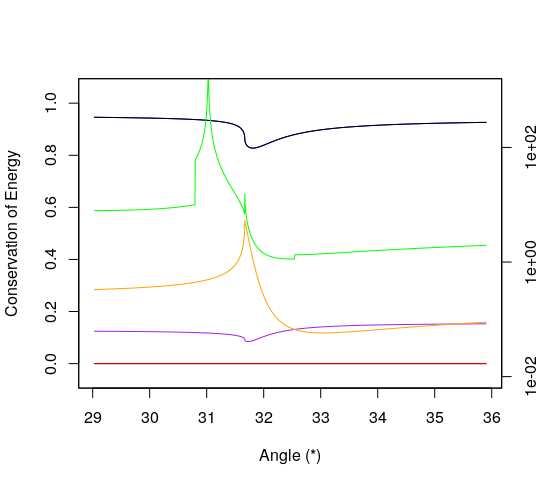}
	\includegraphics[width=7cm]{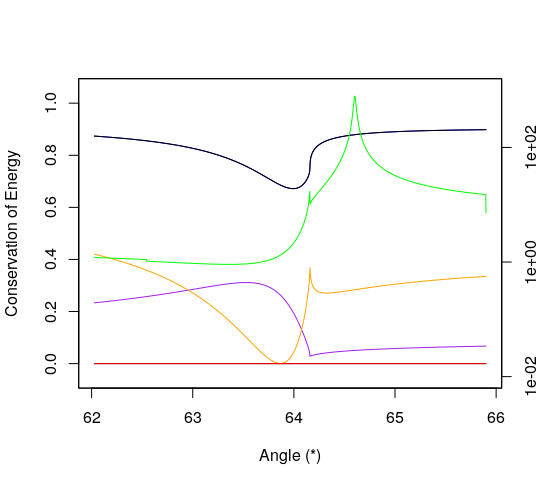}
\centering
	\caption{Conservation of energy spectra for the aforementioned bismuth grating. Two diffraction orders are depicted in orange and purple. Sharp maxima in the C-figure(second axis in green) can be seen to coincide the SPP resonances in the diffracted orders but not the conservation of energy. }	
\centering
\end{figure}

Similarly for antimony two first order SPP’s have been observed which match experimental data at 10.591$\mu$m wavelength(Figure 11). The gratings used were trapezoidal gratings with 50\% duty cycle, profile height of 1.5$\mu$m, period of 20$\mu$m, and a layer depth of 1.4$\mu$m. The SPP’s in this case were not as sharp but were found near the experimentally determined locations of 28.06\degree (C-figure: 27.93\degree) and 36.20\degree(C-figure: 36.06\degree). Although the C-figure presents SPPs which are $<$ \%1 different from the peaks in the diffracted orders, this is a sizable discrepancy for an empirical theory. The trapezoidal geometry of these gratings is not perfectly suited for fast fourier transform approximations, so a test of C-figure maxima vs truncation order was performed.
\begin{figure}[H]
	\includegraphics[width=7cm]{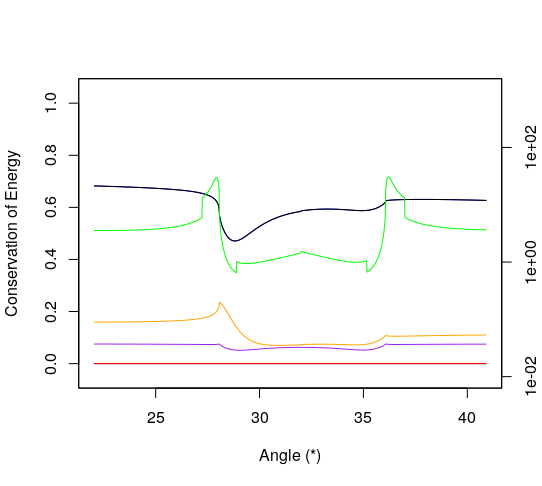}
\centering
	\caption{Conservation of energy spectra for the aforementioned antimony grating. Two diffraction orders are depicted in orange and purple. Sharp maxima in the C-figure(second axis in green) can be seen to coincide the SPP resonances in the diffracted orders but not the conservation of energy. }	
\centering
\end{figure}

It was found that by increasing the truncation order the C-figure slowly approached the locations of resonances in the diffracted orders but did not reach any of them exactly. It is also impractical to use such large truncation orders with fine spectral intervals for mode detection/optimization. More importantly, there are maxima which are attributable to unknown modes in the C-figure that align with extrema in the diffracted orders. Thus, false positives for the identification of SPPs via the C-figure are likely in it's current state. A more rigorous study has been planned to address this issue and adapt the C-figure for better mode distinguishability on corrugated surfaces.

\section*{Conclusion}
Several numerical techniques for the analysis of the planar Fresnel equations and corrugated Chandezon method scattering matrices have been explored. A process for automated planar substrate mode characterization in 3-4 layer arrangements has been demonstrated. Differentiation between surface plasmon resonances and absorbtion losses for periodic corrugated substrates has been achieved with a novel figure of merit and the C-method. Nuances for antimony and bismuth in grating configurations still persist. These nuances suggest that efforts toward method refinement are still required for the figure of merits applicability to the IR spectrum. 

\section*{Acknowledgements}
The authors would like to thank Kofi Edee for their assistance with the C-method operator. Also Stephen Christian for their help with mathematical formalism.


\begin{thebibliography}{14}
\bibitem{NumMethods} 
Dmitry A. Bykov and Leonid L. Doskolovich, "Numerical Methods for Calculating Poles of the Scattering Matrix With Applications in Grating Theory," J. Lightwave Technol. 31, 793-801 (2013) 
\bibitem{ResPoles} 
Didier Felbacq. “Numerical computation of resonance poles in scattering theory” Phys. Rev. E. 64, 047702 – Published 18 September 2001
\bibitem{Roundtrip} 
Jakob Rosenkrantz de Lasson, Philip Trøst Kristensen, Jesper Mørk, and Niels Gregersen, "Roundtrip matrix method for calculating the leaky resonant modes of open nanophotonic structures," J. Opt. Soc. Am. A 31, 2142-2151 (2014)
\bibitem{RCWAvsCmeth} 
N.P. van der Aa. "Diffraction Grating Theory with RCWA or the C Method". Progress in Industrial Mathematics at ECMI 2004. Vol 8 of the series Mathematics in Industry pp 99-103.
\bibitem{LisSMatrix} 
L. Li, Formulation and comparison of two recursive matrix algorithms for modeling layered diffraction gratings, J. Opt. Soc. Am. A 13, 1024–1035 (1996).
\bibitem{Rumpf} 
R. C. Rumpf, "Improved formulation of scattering matrices for semi-analytical methods that is consistent with convention," Progress In Electromagnetics Research B, Vol. 35, 241-261, 2011. 
\bibitem{Nature} 
Jonathan J. Foley, Hayk Harutyunyan, Daniel Rosenmann, Ralu Divan, Gary P. Wiederrecht \& Stephen K. Gray. "When are Surface Plasmon Polaritons Excited in the Kretschmann-Raether Configuration?". Scientific Reports 5, Article number: 9929 (2015) doi:10.1038/srep09929
\bibitem{Book} 
Surface Modes in Physics. Sernelius, B.E. Wiley, 2011.ISBN: 9783527635054. pg 300.
\bibitem{OrigCMeth} 
J Chandezon, G Raoult and D Maystre. A new theoretical method for diffraction gratings and its numerical application. Journal of Optics, Volume 11, Number 4.
\bibitem{AdaptiveSpatial} 
Gérard Granet, Jean Chandezon, Jean-Pierre Plumey, and Karyl Raniriharinosy, "Reformulation of the coordinate transformation method through the concept of adaptive spatial resolution. Application to trapezoidal gratings," J. Opt. Soc. Am. A 18, 2102-2108 (2001) 
\bibitem{DiffFormalism} 
Lifeng Li, Jean Chandezon, Gérard Granet, and Jean-Pierre Plumey, "Rigorous and efficient grating-analysis method made easy for optical engineers," Appl. Opt. 38, 304-313 (1999) 
\bibitem{Cotter} 
N. P. K. Cotter, T. W. Preist, and J. R. Sambles, "Scattering-matrix approach to multilayer diffraction," J. Opt. Soc. Am. A 12, 1097-1103 (1995) 
\bibitem{Bi} 
Khalilzadeh-Rezaie F, Smith CW, Nath J, et al; Infrared surface polaritons on bismuth. J. Nanophoton. 0001;9(1):093792.  doi:10.1117/1.JNP.9.093792.
\bibitem{Sb} 
Justin W. Cleary, Gautam Medhi, Monas Shahzad, Imen Rezadad, Doug Maukonen, Robert E. Peale, Glenn D. Boreman, Sandy Wentzell, and Walter R. Buchwald, "Infrared surface polaritons on antimony," Opt. Express 20, 2693-2705 (2012) 

\end{thebibliography}
\end{document}